\documentclass[square]{ws-procs975x65}

\newcommand{\density}{\varrho}
\newcommand{\factor}{(1-y)[y-x\exp(-x\gamma)]^2}
\newcommand{\fractiongamma}{\left( \frac{2}{5-3\Gamma}\right)%
	^{(5-3\Gamma)/2(\Gamma-1)}}
\newcommand{\modR}{\tilde{R}_0}
\newcommand{\ie}{\textit{i.e.}}
\newcommand{\betapart}[2]{\phi_0 \chi_\infty G^{#1} \pi^2%
	\frac{M^{#2}}{a_\infty^3}}

\begin{document}
\title{Gravastars and bifurcation in quasistationary accretion}
\author{Edward Malec and Krzysztof Roszkowski}
\address{M.~Smoluchowski Institute of Physics, Jagiellonian University,\\
Reymonta 4, 30-059 Krak\'ow, Poland}
\begin{abstract}
We investigate the newtonian stationary accretion of a polytropic perfect fluid
onto a central body with a hard surface. The selfgravitation of the fluid and
its interaction with luminosity is included in the model. We find that for
a given luminosity, asymptotic mass and temperature of the fluid there exist
two solutions with different cores.
\end{abstract}
\bodymatter
\section{Introduction}
The question we want to address in this paper is the following \textit{inverse
problem}: having a complete set of data describing a compact body immersed
in a spherically symmetric accreting fluid, find the mass of the central body.
We assume that we know the total mass, luminosity, asymptotic temperature,
the equation of state of the accreting gas and the gravitational potential
at the surface of the core.

The fundamental question is whether observers can distinguish between
gravastars~\cite{MM} versus black holes as engines of luminous accreting
systems (see a controversy in~\cite{AKL, Narayan}). While we do not address
this problem here, we show a related ambiguity in a simple newtonian model.
\section{The Shakura model}
The first investigation of stationary accretion of spherically symmetric
fluids, including luminosity close to the Eddington limit, was provided by
Shakura~\cite{Shakura}. It was later extended to models including the gas
pressure, its selfgravity and relativistic effects~\cite{OS, Thorne, RM, Park}.

In the following we will denote the areal velocity by~$U(r,t)=\partial_t R$
(where~$t$ is comoving time and $R$ the areal radius), the local, Eddington and
total luminosities by~$L(R)$, $L_E$ and~$L_0$, quasilocal mass by~$m(R)$ and
total by~$M$, pressure by~$p$, the baryonic mass density by~$\density$ (the
polytropic equation of state will be $p = K \density^\Gamma$,
$1< \Gamma \leq 5/3$) and the gravitational potential by~$\phi(R)$. The radius
of the central body is~$R_0$ while its ``modified radius'' is defined by
$\modR = GM/|\phi(R_0)|$. Under the assumption that at the outer boundary of
the fluid the following holds true:
\begin{equation}
U^2_\infty \ll \frac{G m(R_\infty)}{R_\infty} \ll a^2_\infty,
\end{equation}
we have the following set of equations:
\begin{equation}
\dot M = - 4 \pi R^2 \density U,
\label{roszkowski:eq1}
\end{equation}
\begin{equation}
U \partial_R U = -\frac{G m(R)}{R^2} - \frac{1}{\density} \partial_R p
+ \alpha \frac{L(R)}{R^2},
\end{equation}
\begin{equation}
\partial_R \dot M = 0,
\end{equation}
\begin{equation}
L_0 - L(R) = \dot M \left( \frac{a^2_\infty}{\Gamma - 1}
- \frac{a^2}{\Gamma -1} - \frac{U^2}{2} - \phi(R) \right).
\label{roszkowski:eq4}
\end{equation}
$\alpha$ is a dimensional constant $\alpha = \sigma_T / 4 \pi m_p c$.
The details of solving the system are provided elsewhere~\cite{MRK, MalecR}
and we will present only the main results here.

We assume that the accretion is critical, \ie, there exists a sonic point,
where the speed of accreting gas~$U$ is equal to the speed of sound~$a$.
All values measured at that point will be denoted with an asterisk. We define:
\[
x = \frac{L_0}{L_E}, \qquad y = \frac{M_\ast}{M}, \qquad
\gamma = \frac{\modR}{R_\ast},
\]
and obtain the total luminosity:
\begin{equation}
L_0 = \betapart{2}{3} \factor \fractiongamma .
\label{roszkowski:luminosity}
\end{equation}
$\chi_\infty$ is approximately the inverse of the volume of the gas located
outside the sonic point. In sake of brevity we will use
\[
\beta = \alpha \betapart{}{2} \fractiongamma
\]
to obtain \eref{roszkowski:luminosity} in a form using the relative luminosity:
\begin{equation}
x = \beta \factor .
\label{roszkowski:relative}
\end{equation}
\section{Bifurcation}
For the relative luminosity, fulfilling~\eref{roszkowski:relative} we proved
the following theorem:
\begin{romanlist}[(iii)]
\item For the functional $F(x, y) = x - \beta \factor$ there exists a critical
point $x =a$, $y=b$ such that $F(a, b) = 0$ and $\partial_y F(a, b) = 0$ with
$0 < a < b < 1$ and $a = 4 \beta (1-b)^3$, $b = [2+a \exp(-a \gamma)]/3$.
\item For any $0 < x < a$ there exist two solutions $y(x)^+_-$ bifurcating from
$(a,b)$. They are locally approximated by:
\begin{equation}
y^+_- = b \pm \left(\frac{(a-x)[b+a\exp(-a\gamma)(1-2a\gamma)]}%
{\beta[b-a\exp(-a\gamma)][1-a\exp(-a\gamma)]} \right)^{1/2} .
\end{equation}
\item The relative luminosity~$x$ is extremized at the critical point~$(a,b)$.
\end{romanlist}
\section{Discussion}
In the paper we have assumed the existence of an accreting system which
satisfies certain conditions. Under those assumptions the complicated set of
integro-differential nonlinear equations~(\ref{roszkowski:eq1}--%
\ref{roszkowski:eq4}) can be simplified to an algebraic
one~(\ref{roszkowski:relative}). We checked numerically that the performed
simplification causes errors of the order of $10^{-3}$ (see~\cite{MRK}
for details).

The analysis of \eref{roszkowski:relative} shows that there exist two different
solutions, having the same total luminosity and total mass, but different masses
of the core objects. One can also conclude that for sufficiently large~$\beta$
the maximal relative luminosity~$a$ can get close to~$1$, \ie, the total
luminosity approaches the Eddington limit.

As the two solution branches bifurcate from the point $(a, b)$, there is no
much difference between the central masses of bright objects (see~\cite{MalecR,
MRK} for plots). However, when luminosity is small ($L_0 \ll L_E$), this
difference can become arbitrarily large. This can be understood intuitively,
because the radiation is small for test fluids (since the layer of gas is thin),
or when the central object is light (therefore weakly attracting surrounding
gas).

The results obtained here are consistent with relativistic analysis neglecting
interaction between the gas and the radiation~\cite{KKMMS, KMM}.
\section*{Acknowledgments}
This paper has been partially supported by the MNII grant 1P03B 01229.

\end{document}